# Quantifying Local Strain Field and Deformation in Active Contraction of Bladder Using a Pretrained Transformer Model: A Speckle-Free Approach

Authors:

Alireza Asadbeygi (1), Anne Robertson (1,2), Yasutaka Tobe (1),

Masoud Zamani (1), Sean D. Stocker (3), Paul Watton (4), Naoki Yoshimura (5,6), Simon C Watkins (7)

[1]Department of Mechanical Engineering and Materials Science, University of Pittsburgh, PA, U.S.A.

[2]Department of Bioengineering, University of Pittsburgh, PA, U.S.A.

[3]Department of Neurobiology, University of Pittsburgh School of Medicine, PA, U.S.A.

[4]Department of Computer Science & Insigneo Institute for in silico Medicine, University of Sheffield, Sheffield, U.K.

[5]Department of Pharmacology and Chemical Biology, University of Pittsburgh, Pittsburgh, PA, U.S.A.

[6]Department of Urology, University of Pittsburgh, Pittsburgh, PA, U.S.A.

[7]Center for Biologic Imaging, University of Pittsburgh, Pittsburgh, PA, U.S.A.

Highlights:

1- Introduced and validated a speckle-free strain mapping method for large deformation of soft tissues
2- Developed isotonic biaxial device compatible with multiphoton microscopy.
3- Quantified anisotropic strain fields during active bladder contraction

## Abstract

Accurate quantification of local strain fields during bladder contraction is essential for understanding the biomechanics of bladder micturition, in both health and disease. Conventional digital image correlation (DIC) methods have been successfully applied to various biological tissues; however, this approach requires artificial speckling, which can alter both passive and active properties of the tissue. In this study, we introduce a speckle-free framework for quantifying local strain fields using a state-of-the-art, zero-shot transformer model, CoTracker3. We utilized a custom-designed, portable isotonic biaxial apparatus compatible with multiphoton microscopy (MPM), to demonstrate this approach, successfully tracking natural bladder lumen textures without artificial markers. Benchmark tests validated the method's high pixel accuracy and low strain errors. Our framework effectively captured heterogeneous deformation patterns, despite complex folding and buckling, which conventional, DIC often fails to track. Application to in vitro active bladder contractions in four rat specimens (n=4) revealed statistically significant anisotropy (p<0.01), with higher contraction longitudinally compared to circumferentially. Multiphoton microscopy further illustrated and confirmed heterogeneous morphological changes, such as large fold formation during active contraction. This non-invasive approach eliminates speckle induced artifacts, enabling more physiologically relevant measurements, and has broad applicability for material testing of other biological and engineered systems.

## 1.Introduction

Accurate measurement of tissue local strain is essential for understanding the biomechanical behavior of soft tissues, for both active and passive functions. In bladder tissue, quantifying strain fields is essential, not only for elucidating the mechanisms of normal filling and voiding functions, but also for assessing pathological conditions such as bladder outlet obstruction (BOO) and urinary incontinence. For instance, prior research has demonstrated that bladder obstruction affects contractile behaviors in both human and animal models [1-5]. In a recent human subject study, Choi et al. [2] reported that in individuals without bladder outlet obstruction (BOO), age showed a significant association with bladder contractility (p=0.021), whereas this relationship was not observed in those with BOO (p=0.057). These findings suggest that BOO may serve as a confounding factor when evaluating age-related changes in bladder contractility.

Conventional 2D histology and scanning multiphoton microscopy have laid the essential groundwork for our current understanding of bladder wall structure. Seminal studies based on fixed-tissue histology using scanning electron microscopy (SEM) [6-8], confocal microscopy [6, 9-12], and multiphoton microscopy [13-16] of unfixed specimens have exquisitely characterized the layered architecture of the bladder wall, revealing critical insights into smooth muscle organization, extracellular matrix composition, and cellular morphology. Building on this structural knowledge, extensive mechanical testing has characterized the passive properties of the bladder wall and its constituent layers [14, 15, 17-22]. For example, Cheng et al. [15] used a custom biaxial testing system combined with multiphoton microscopy to quantify layer-dependent collagen recruitment during bladder loading, demonstrating how wall extensibility arises from the coordinated engagement of the lamina propria and detrusor layers. Similarly, Hornsby et al. [21] investigated the functional morphology of the murine bladder wall under uniaxial loading using multiphoton microscopy and related collagen fiber organization to mechanical response, confirming that low-strain deformation is dominated by unfolding of rugae in the lamina propria. These and other passive studies have established that the compliant behavior of the bladder wall depends critically on collagen architecture and depth-dependent mechanics across tissue layers.

In contrast, active contraction studies have primarily focused on regional differences in the bladder response [23-26] or the effect of cellular mechanisms and pharmacological agents on bladder contractility [4, 27-30]. While valuable, these studies have not considered the local nature of these spatial deformations. For instance, Trostorf et al. [31] utilized ex vivo porcine bladders to examine active behavior under various pressure conditions, providing valuable insights into pressure–volume relationships but without resolving local strain fields. Likewise, Borsdorf et al. [24] investigated the location-dependency of active properties in porcine bladder strips in isometric contraction tests, showing that maximum active stress was higher in the longitudinal direction compared to the circumferential direction. However, despite these advances, there remains a lack of spatially resolved data describing how the bladder wall deforms locally during active contraction.

Conventional digital image correlation (DIC) methods have been widely employed for strain measurement in both engineering and biomedical applications [32-36]. Traditional DIC and optical measurements have been used to characterize the heterogenous properties of cerebral [32] and aortic aneurysms [37, 38]. As an example, Shih, E.D. et al. [32] introduced a hybrid DIC-Finite Element approach to quantify the anisotropy and heterogeneity in cerebral aneurysm tissue during passive stretch-controlled mechanical testing, and correlated this response with tissue structure such as collagen fiber organization. However, the application of DIC to biological tissues, especially during active contraction, is hindered by several challenges. Primarily, DIC typically requires the creation of a speckle pattern on the tissue surface. This process necessitates removing the specimen from oxygenated Krebs solution for extended periods, causing surface drying that can diminish smooth muscle contractility. In addition, the inks used for speckling may coat or penetrate the tissue surface, altering its mechanical properties [31, 39, 40]. Second, the reliance of DIC on consistent lighting, precise camera alignment, speckle pattern quality, and image resolution makes it vulnerable to errors in complex deformations [41-43]. Recent studies have investigated the short comings of the conventional DIC approaches for large deformations [44].

Recent advances in deep learning have introduced alternative approaches for conventional DIC strain calculation [45-48]. For instance, Yang, R. et al. [46] presented a deep learning based strain measurement method, by training a convolutional neural network architecture on synthetic speckled images. While achieving high accuracy on benchmark cases (max error: 0.083 pixels), the model was applied to a synthetic training dataset with relatively small deformations (max deformation of 5 pixels), and dependent on speckled surfaces. In a more recent study, Y. Yang et al [44] proposed a new deep learning based DIC approach (DICnet3+) in which they strove to address the short comings of the earlier models for large deformations, by diversifying the training data to include maximum deformations up to 25 pixels. By utilizing Convolutional Neural Networks (CNNs), these architectures can extract deep features from each pixel for correlation, offering greater robustness than traditional DIC methods that rely primarily on raw pixel intensities [44].

However, these methods have generally been trained on extensive datasets of synthetic speckle patterns or trained on modality-specific datasets (e.g. ultrasound [49, 50]). To our knowledge, no speckle-free deep learning approach has been successfully tested to enable zero-shot (no fine tuning) calculation of strain fields in biological tissues, particularly for the large and complex deformations associated with active contraction.

Despite these advances, visualizing and quantifying bladder deformation in three dimensions during active contraction remains a significant challenge. In this study, we address this gap by developing a custom portable biaxial isotonic apparatus compatible with multiphoton microscopy (MPM) to allow for isotonic contraction under imaging. Furthermore, we introduce a speckle-free strain-calculation framework that employs CoTracker3 (particle-tracking model) to quantify the strain field during active contraction. This multimodal approach not only preserves tissue viability by eliminating artificial speckling but also provides insight into the heterogeneous deformation patterns intrinsic to active bladder contraction.

## 2. Methods

The methods section is organized into five main parts. First (Section 2.1), we describe the design and fabrication of a custom, portable isotonic biaxial contraction apparatus compatible with multiphoton microscopy. Second (Sections 2.2), we detail two complementary imaging modalities used to capture deformation: multiphoton microscopy (MPM) for high-resolution surface topology before and after contraction, and high-magnification video microscopy for dynamic tracking of tissue motion during active contraction. Third (Section 2.3), we present the deep learning–based strain calculation framework. Fourth (Section 2.4), we describe the benchmark tests used to validate the framework against analytical solutions and measurements, followed by Section 2.5, which details the metrics used for the accuracy evaluation of these benchmarks. Finally, in Section 2.6, we apply the framework to bladder contraction videos to calculate and analyze the resulting strain fields.

### 2.1. Experimental Setup:

#### 2.1.1. Biaxial Isotonic Contraction Apparatus:

A critical design requirement for the biaxial apparatus was compatibility with the multiphoton microscope (Nikon A1R MP HD, Tokyo, Japan), particularly regarding chamber size and lens operational constraints. Additionally, the tissue must be securely clamped without damage and kept viable for active contraction throughout long scan times (5 to 10 hours). Commercial isotonic systems typically do not fit under multiphoton microscopes.

Inspired by the uniaxial isotonic setups in previous studies[51], a custom biaxial isotonic apparatus was developed to test bladder tissue specimens under physiological conditions. The system includes a temperature-controlled tank filled with KREBS solution and maintained at 37 °C by a digital controller and a base heater. Fixtures, mounts, and rake holders were fabricated by 3D printing polylactic acid (PLA). To minimize the effects of friction and the weight of the rakes on the applied isotonic load, the rake holders were designed to be 90% porous, allowing them to float within the testing tank.

To accommodate mounting of the small (~5 mm wide) and relatively thick (1mm) bladder specimen, rakes were custom-manufactured with stainless steel needles with diameters of 250 µm. This spacing allowed positioning of three rakes across the specimen width. Each claw measured 3 mm in length to reliably engage bladder walls up to 1–2 mm thick in the contracted state. The biaxial loading was applied isotonically using weights suspended from a triple-pulley mechanism (Figure 1a), providing consistent and adjustable tension during the tests.

### 2.1.2. Sample Preparation and Loading Protocol:

Bladders were harvested from 10-week-old male Sprague Dawley (SD) rats. The bladder was cut open longitudinally, trimmed into a square shape (~5 mm × 5 mm), and mounted on the custom rakes using a 3D-printed platform (Figure 1b) similar to previously published works [33]. The tissue was oriented such that the loading axes aligned with the longitudinal and circumferential anatomical directions of the bladder. The tank was filled with KREBS solution [33] aerated by 95% $O_2$ and 5% $CO_2$. A weight of 1 g (or 0.5 g for video tracking) was applied in each of the two biaxial directions. These loads were selected based on pilot studies to maintain tissue tautness for planar imaging without inducing excessive pre-stretch that might inhibit active contraction. Active contraction was induced by replacing the initial Krebs solution with a homogenized high-concentration KCl solution (80 mM) [52, 53].

### 2.2. Imaging Modalities

### 2.2.1. Multiphoton Microscopy (MPM) of Bladder Topology Before and After Contraction:

Two MPM scans were performed (one before and one after contraction) to capture surface topology. Imaging parameters were optimized to shorten scan duration and maintain viability (Nikon 10x objective, S Fluor 0.50NA DIC N1, 4µm Z-step, resonant scan with 1.24 µm/pixel resolution). To prevent image distortion from sample motion induced by movement of the microscope stage, the bladder was stabilized with a cover glass placed over the bladder lumen. The resolution of scans was enhanced using GPU rendering capabilities of IMARIS (Oxford Instruments plc, UK) and MPM image stacks were then 3D-reconstructed based on the exported surface files in IMARIS. The rendered video of the 3D MPM scan of the contracted bladder is provided in supplementary materials.

To visualize the lumen surface topology and layer-wise wall structure in more detail, a second bladder was obtained from a 10-week-old male SD rat and subjected to the same isotonic contraction protocol. After contraction, the KREBS solution was replaced with 4% paraformaldehyde (PFA) to fix the tissue in the contracted state. A strip specimen was cut from the central region and stained for α-smooth muscle actin (αSMA), following prior protocols [16, 54]. Briefly, the fixed sample was incubated with monoclonal mouse anti-human smooth muscle actin clone 1A4 IgG2a antibody (Dako, Denmark) as the primary antibody and Alexa Fluor 568 goat anti-mouse IgG2a (Invitrogen, USA) for the secondary antibody. MPM imaging was subsequently performed on cross-sections of the strip at high magnification, using a Nikon APO LWD 25x water immersion objective lens (NA 1.10), yielding a spatial resolution of 0.49 µm/pixel in the resonant scanning mode.

### 2.2.2. High Magnification Video Capturing of Active Contraction

For dynamic strain analysis, bladders (n = 4) were harvested from 10-week-old SD male rats, cut open longitudinally, trimmed, and mounted similarly to Section 2.1.2. Active contraction was induced by high-concentration KCl, as in Section 2.1.2. To enable larger contractile deformations, the isotonic load was reduced to 0.5 g for this part of the study. High-magnification videos were then captured from the bladder lumen during active contraction, using a cost-effective digital microscope (<$200) (TOMLOV DM601, China).

## 2.3. Strain Calculation Framework

A novel deep learning-based framework was developed to determine the local strain field. This framework is comprised of two main modules:

### 2.3.1. Tracking Virtual Grid Points (CoTracker3)

The state-of-the-art deep learning particle tracking model CoTracker3 [55] was used to track the virtual points on the samples, based on the bladder lumen texture and topology, without the need for speckling. Crucially, we employed this pre-trained model in a "zero-shot" manner, meaning no fine-tuning or training on our specific bladder tissues was required. The CoTracker3 model is a sophisticated point tracking system that leverages image features to accurately follow points across frames. It utilizes a transformer-based architecture, which excels at capturing long-range dependencies and contextual information within the image sequences [55].

### 2.3.2. Calculating The Strain Field Based on the Trajectories of the Tracked Points:

Using the point trajectories from CoTracker3, we computed the strain tensor at various field-of-view locations. First, the structured grid points in the initial configuration were triangulated to create a computational mesh, using the Delaunay method in Python SciPy library. The displacement vector $\boldsymbol{u}(\boldsymbol{x},\boldsymbol{t})$ was determined by tracking the motion of each grid point from its initial position to its deformed position $\boldsymbol{x}(\boldsymbol{t})$.

To calculate the two-dimensional strain field, we first computed spatial derivatives of $\boldsymbol{u}(\boldsymbol{x},\boldsymbol{t})$ and used this to calculate the deformation gradient tensor $\boldsymbol{F}$ and the Green–Lagrange strain tensor $\boldsymbol{E}$,

$$\boldsymbol{F} = \boldsymbol{\nabla u} + \boldsymbol{I} \qquad (1)\,[56]$$

$$\boldsymbol{E} = \frac{1}{2}(\boldsymbol{F}^T\boldsymbol{F} - \boldsymbol{I}) \qquad (2)\,[56]$$

The strain tensor is approximated across each triangular element, and linear triangular shape functions are subsequently employed to derive the nodal strain components, similar to the previous works [57, 58]. Finally, contour plots are generated for each strain component.

## 2.4. Benchmark and Validation Tests:

Five benchmark tests were designed to evaluate the performance of the strain calculation framework, progressing from theoretical validation to physical testing and finally to biological texture verification.

### 2.4.1. Synthetic Benchmarks (Validation against Analytical Solutions)

Synthetic animations were generated for three fundamental deformation modes using synthetic speckle patterns. The frames of these speckle images were created with the synthetic speckle generation function in the µDIC Python library [59], and the videos were assembled from these frames using the imageio Python library.

1. **Equibiaxial stretch**: The deformation for 2D equibiaxial stretch is defined by:

$$x(X,Y) = \gamma X \quad (3), \quad y(X,X) = \gamma Y \quad (4)$$

In which the $X$ and $Y$ are the coordinates in the reference configuration and $\gamma$ is the stretch parameter (here $\gamma = 1.5$). From equations 1 and 2, the corresponding deformation gradient and strain tensors are:

$$F = \begin{bmatrix} \frac{\partial x}{\partial X} & \frac{\partial x}{\partial Y} \\ \frac{\partial y}{\partial x} & \frac{\partial y}{\partial Y} \end{bmatrix} = \begin{bmatrix} \gamma & 0 \\ 0 & \gamma \end{bmatrix} = \begin{bmatrix} 1.5 & 0 \\ 0 & 1.5 \end{bmatrix} \quad (5), \quad E = \begin{bmatrix} 0.625 & 0 \\ 0 & 0.625 \end{bmatrix} \quad (6)$$

2. **Rigid body rotation:** The imposed 2D rigid body motion is defined by

$$x(X,Y) = X\cos\theta - Y\sin\theta \quad (7), \quad y(X,Y) = X\sin\theta + Y\cos\theta \quad (8)$$

so that the deformation gradient

$$F = \begin{bmatrix} \cos\theta & -\sin\theta \\ \sin\theta & \cos\theta \end{bmatrix} \quad (9)$$

is pure rotation. Since the rotation matrices are orthogonal, the Green–Lagrange strain tensor is identically zero.

3. **2D shear deformation:** the 2D shear deformation, is defined as:

$$x(X,Y) = X + \gamma Y \quad (10), \quad y(X,Y) = Y \quad (11)$$

For the special case $\gamma = 0.5$, the deformation gradient and strain tensors are calculated based on equations 1 and 2:

$$F = \begin{bmatrix} \frac{\partial x}{\partial X} & \frac{\partial x}{\partial Y} \\ \frac{\partial y}{\partial x} & \frac{\partial y}{\partial Y} \end{bmatrix} = \begin{bmatrix} 1 & \gamma \\ 0 & 1 \end{bmatrix} = \begin{bmatrix} 1 & 0.5 \\ 0 & 1 \end{bmatrix} \quad (12), \quad E = \begin{bmatrix} 0 & 0.25 \\ 0.25 & 0.125 \end{bmatrix} \quad (13)$$

To evaluate the model's performance, the tracking error of the grid points and the strain error were compared against the results from analytical solutions.

### 2.4.2 Experimental Data: Uniaxial Tensile Test of a Dog-Bone-Shaped Latex Sample:

To demonstrate the framework's capability on physical mechanical testing, a dog-bone-shaped latex sample was cut using a custom-made cutter [60]. Because the surface of the latex had no inherent texture, it was speckled using an airbrush and black ink to create a minimal texture pattern. The sample was then stretched uniaxially in a stretch-controlled tensile apparatus to a stretch of 1.27 (from clamp to clamp). A digital microscope recorded the deformation process, and the resulting video was processed by the framework to estimate the local strain field. To evaluate the accuracy in this case, random manual measurements of strain (based on speckles) were made using Fiji [61] and compared to the corresponding values predicted by the framework.

### 2.4.3. Synthetic Deformation on Real Bladder Tissue

To assess the framework's performance on actual biological texture (rather than synthetic speckles or inked latex) before applying it to the experimental data, we created a final benchmark case using an image of a bladder sample (Sample 1). A synthetic equibiaxial contraction (λ= 0.6) was applied to the image mathematically using μDIC library. The framework was then tested on this deformation to verify if it could accurately recover the known ground-truth strain from the natural, texture-sparse bladder surface.

### 2.5. Accuracy Evaluation

To assess the performance of the proposed tracking and strain-estimation framework, we quantified both pixel-level tracking accuracy and strain field reconstruction consistency across all deformation scenarios. For each experiment, the displacement and strain fields predicted by CoTracker3 were compared against corresponding ground-truth solutions obtained either analytically (for synthetic deformations) or from imposed deformation fields.

The tracking accuracy was measured using the root-mean-square error (RMSE) between predicted and true pixel trajectories. Similarly, strain accuracy and spatial consistency were evaluated using two distinct metrics based on the signed error at each grid point $i$, defined as $e_i = E_{pred,i} - E_{true,i}$.

First, the Root Mean Square Error (RMSE) was used to quantify the global accuracy and magnitude of the error across $N$ grid points:

$$RMSE = \sqrt{\frac{1}{N} \sum_{i=1}^{N} e_i^2} \qquad (14)$$

Second, the Spatial Standard Deviation (SD) was used to evaluate the precision and spatial consistency of the error distribution across the mesh:

$$SD = \sqrt{\frac{1}{N} \sum_{i=1}^{N} (e_i - \bar{e})^2} \qquad (15)$$

where $\bar{e}$ is the mean signed error.

By reporting these metrics separately, we distinguish between the model's overall accuracy (RMSE) and its ability to maintain a uniform error distribution (SD). For visualization, we generated contour plots of the relative strain error (%) for each strain component. For components with a ground-truth value of zero (e.g., $E_{i,j}$ in rigid rotation), the absolute error was plotted instead.

### 2.6. Applications of Framework to Bladder Tissue Under Active Contraction

The videos recorded during bladder contraction (Section 2.2.2) were processed using the developed framework. Here, the natural features of the bladder surface were used for point tracking, without any speckling steps. A 7x7 tracking grid was initialized on the relaxed tissue (reference configuration). The grid density was selected based on the texture sparsity of the bladder surface and the size of the folds appearing during the contraction, ensuring that tracking points were not occluded within the folds. Following strain calculations, a paired t-test was conducted between the strain components at each node to evaluate anisotropy in the contractile strains.

## 3. Results

### 3.1. Multiphoton Microscopy of Contracted Bladder: 3D Scanning MPM and Bladder Cross Section Imaging Reveals Changes in Surface Topology and Creation of Large Folds After Contraction.

#### 3.1.1. Comparison of Lumen Surface Topology Before and After Contraction

Figures 2a and 2b shows projection of stacks of MPM images from the lumen side of the bladder before and after contraction, respectively, while under isotonic loading. The red and green channels correspond to collagen and urothelial cells, respectively. The crosshair lines show the location of the in-silico bisection that is visible on the right and bottom of the MPM images. Before contraction, small "rugae" with an amplitude of ~100 µm are visible in Figure 2a. These rugae are largely oriented longitudinally (vertical in Figure 2), resulting in a relatively flat surface. During contraction, the lamina propria buckles and large folds form on the bladder lumen. These folds create alternating convex and concave surfaces, reaching a maximum height of ~900 µm. The bladder thus undergoes heterogeneous deformation. The folds exhibit greater magnitude and longer wavelength in the upper region, near the bladder dome, than in the lower region, near the trigonal area. These large folds are also visible in the vertical and horizontal bisections (white arrow). Although the urothelial cells cover the surface of the lumen, they are folded together in convex areas, resulting in a higher intensity green signal. (A video rendering of the 3D MPM scan of the lumen is provided in the supplementary materials.)

#### 3.1.2. Cross-Sectional View of SMC And Collagen Fiber Architecture in The Bladder Wall

The sample prepared for the cross-sectional scan is shown in Figure 3c. Figure 4 presents an orthogonal cross section through the bladder's central region. Autofluorescent collagen fibers appear in red, while αSMA-stained smooth muscle cells (SMCs) appear in orange. The multilayered structure of the bladder wall is clearly visible. Consistent with the lumen-side MPM images (Figure 2), prominent folds in the lamina propria (top layer) are observed. The SMC bundles exhibit heterogeneous, multidirectional orientations, with primary alignments in the longitudinal and circumferential directions (Figures 4b and 4c). Interactions between collagen fibers and SMC bundles, as well as the presence of intramural collagen fibers, are also visible.

### 3-2- Model Validation on Benchmark Tests:

To evaluate the accuracy of our strain-calculation framework, we report two primary error metrics in Table (1): (1) pixel-level tracking error and (2) strain-component error. All mean error values are reported as Root Mean Square Error (RMSE) with Standard Deviation (SD).

Table 1. Summary of Tracking and Strain Calculation Errors for Benchmark Tests

| Benchmark Test | Tracking Error (pixels) | Strain Error |
|---|---|---|
| Equibiaxial | RMSE=1.31, SD= 0.62 | Exx: RMSE=0.0073  SD=0.0053 ; Eyy: RMSE=0.0123  SD= 0.0109;   Exy: RMSE=0.0047  SD= 0.004 |
| Rigid Body Rotation | RMSE=1.29, SD= 0.64 | Exx: RMSE= 0.0033  SD= 0.0031 ; Eyy: RMSE=0.0045  SD= 0.0043;   Exy: RMSE=0.0037  SD= 0.0035 |
| 2D Shear | RMSE=1.21, SD=0.59 | Exx: RMSE=0.0058 SD= 0.0051 ; Eyy: RMSE=0.0063  SD=0.0061;   Exy: RMSE=0.0045  SD= 0.0041 |
| Latex Sample | - | Exx: RMSE=0.0079 SD= 0.005 ;   Eyy: RMSE=0.0130  SD= 0.010 |
| Synthetic Bladder Contraction | RMSE=1.48, SD=0.81 | Exx: RMSE=0.0052  SD= 0.0049 ; Eyy: RMSE=0.0067  SD= 0.0056;   Exy: RMSE=0.0042  SD=0.0039 |

### 3.2.1. Analytical Benchmarks (Synthetic Speckles)

- **Equibiaxial Stretch:** The contour maps of strain errors are shown in Figure 5. Under very large deformation (maximum displacement ≈ 140 pixels), the model achieved strong tracking performance with an RMSE of 1.31 pixels (spatial SD: 0.62 pixels**).** The analytical solution predicts equal normal strains of $E_{xx} = E_{yy} = 0.625$. The reconstructed strain components showed excellent agreement with the analytical values, with absolute errors of

  $E_{xx}$: RMSE = 0.0073, SD = 0.0053, $E_{yy}$: RMSE = 0.0123, SD = 0.0109.

  The shear strain component, analytically zero, was also accurately recovered with

  $E_{xy}$: RMSE = 0.0047, SD = 0.0040,
  indicating model accuracy for zero-strain modes.

- **Rigid Body Rotation:** For rigid rotation, the analytical strain field is identically zero. The contour maps of absolute error are shown in Figure 6. The framework successfully preserved this condition, producing very small absolute strain errors:

  $E_{xx}$: RMSE = 0.0033, SD = 0.0031, $E_{yy}$: RMSE = 0.0045, SD = 0.0043 and $E_{xy}$: RMSE = 0.0037, SD = 0.0035.

  Tracking accuracy remained high, with an RMSE of 1.29 pixels (SD: 0.64 pixels) with a maximum displacement of 105 pixels, confirming the model's ability to distinguish rigid motion from deformation.

- **2D Shear Deformation:** The contour maps of strain errors are presented in Figure 7. In the shear deformation benchmark, the analytical strain field contains a dominant shear component and small normal strains. The model accurately captured both:

  $E_{xy}$: RMSE = 0.0045, SD = 0.0041, $E_{yy}$: RMSE = 0.0063, SD = 0.0061, $E_{xx}$: RMSE = 0.0058, SD = 0.0051

This case also produced the lowest tracking error among the analytical benchmarks, with RMSE = 1.21 pixels (SD: 0.59 pixels) with a maximum displacement of 200pixels, reflecting stable performance under asymmetric deformation.

### 3.2.2. Validation on Uniaxial Tensile Test (Latex)

The tracked grid and the resulting strain contours from the framework are shown in Figure 8. At the right and left ends of the region of interest, the local strain values are close to the clamp-to-clamp strain ($E_{xx}$). As expected, the maximum absolute values of the $E_{xx}$ (0.39) and $E_{yy}$ (0.119) were observed in the middle region due to the smaller cross sectional area. To assess the accuracy of the model prediction in this case, five random pairs of speckle particles were chosen, and the local stretch was measured manually using Fiji. Comparison of the predicted strain fields (Figure 8) with manual strain measurements (n=5 random pairs) indicated a RSME of 0.013 (SD=0.010) for $E_{yy}$ and 0.0079 (SD=0.005) for $E_{xx}$, validating the framework's capability in real physical testing environments.

### 3.2.3. Synthetic Deformation on Real Bladder Tissue

To evaluate the framework on natural bladder texture, a synthetic equibiaxial contraction ($\lambda = 0.6$) was applied to a real bladder image (Section 2.4.3). Under this imposed deformation, the analytical strain field yields $E_{xx} = E_{yy} = -0.32$. Despite the low contrast of the bladder lumen in grayscale and downsampling of the image to 800 × 800 pixels, the model achieved a tracking RMSE of 1.48 pixels (spatial SD: 0.81 pixels). The reconstructed strain components showed low absolute error relative to the analytical field:

$E_{xx}$: RMSE = 0.0052, SD = 0.0049, $E_{yy}$: RMSE = 0.0067, SD = 0.0056.

Contour maps of the relative error for strain-component ($E_{xx}, E_{yy}$) are shown in Figure 9, demonstrating the error distribution and confirming the stability of the method when applied to natural tissue texture. These results verify that the speckle-free, zero-shot approach maintains high accuracy on large contractile deformation even on biological images lacking artificial texture.

### 3.3. Application of Framework to Quantify Strain During Bladder Contraction:

The tracking grid and the strain contours for four bladder samples undergoing active contraction are shown in Figures 10 to 13. As for the validation cases, the grid points were initialized in the first frame of the video, prior to contraction, and subsequently tracked throughout the contraction process. The frame in which the tissue stopped contractile deformation and reached equilibrium was used as the final configuration to calculate the strains (Strain contour plots for sample 2, at multiple time points during the contraction process are provided in the supplementary materials). As qualitatively observed, the model successfully tracked the initial grid points based solely on the bladder lumen texture, without any artificial speckling. Consistent with the multiphoton microscopy (MPM) images (Figure 2), the small surface rugae transformed into larger folds during contraction. As expected, all strain components exhibited negative values, indicating tissue contraction. In all four samples, the strain contours appeared heterogeneous, and the components $E_{xx}$ and $E_{yy}$ displayed distinct spatial distributions with different strain concentration zones. Qualitatively, the strain in the longitudinal direction of the bladder ($E_{yy}$) demonstrated greater contraction compared to the circumferential direction ($E_{xx}$).

A summary of the temporal evolution of mean strain components, up to the contraction endpoint, is shown in Figure 14. Both $E_{xx}$ and $E_{yy}$ magnitudes increased monotonically with time as contraction progressed, but the magnitude of $E_{yy}$ consistently exceeded that of $E_{xx}$ across all samples. This observation demonstrates that bladder tissue undergoes greater longitudinal shortening relative to the circumferential direction, highlighting an anisotropy in the contractile response. Error bars indicate the spatial standard deviation of each component, reflecting the heterogeneous nature of the deformation.

Violin plots of these two components are presented in Figure 15. To quantitatively assess the anisotropy in contractile strains, paired t-tests were performed on nodal strain component values (N = 49), with resulting p-values provided in the violin plots. As shown, $E_{yy}$ was statistically significantly larger than $E_{xx}$ in all three samples (p-values < 0.01).

## 4- Discussion

In this study, we introduced a speckle-free, zero-shot deep learning framework leveraging the transformer-based CoTracker3 model for quantification of heterogeneous strain fields during active bladder contraction. Our approach effectively tracked virtual grid points directly from natural bladder surface textures, eliminating the conventional need for artificial speckling, which can compromise tissue viability and alter both passive and active mechanical properties. Through comprehensive benchmark tests (including synthetic equibiaxial stretching, rigid body rotation, shear deformation, and a physical uniaxial tension test) we validated the robustness and accuracy of our framework. Computed strain fields consistently exhibited strong agreement with both analytical predictions and manual measurements, demonstrating the accuracy and practical applicability of the proposed method.

Notably, our method enabled, for the first time, detailed visualization and accurate quantification of strain distributions during active bladder contractions without relying on specialized speckling techniques, complex camera settings or extensive model training. This contrasts with conventional Digital Image Correlation (DIC) methods, which often fail under large or complex deformations (such as contraction-induced folding) because speckle patterns may tear or decorrelate, leading to boundary instability and erroneous data [44]. In the case of bladder contraction, such discontinuities are particularly problematic since speckles can vanish between folds, impairing displacement tracking accuracy.

Recent supervised deep-learning DIC frameworks (e.g., UNet3+, DICNet, and DICNet3+) have reported very low tracking RMSE values on the order of 0.2–0.5 pixels on speckled benchmark datasets with relatively smaller displacements than our cases. In comparison, our CoTracker3-based framework achieved tracking RMSE values of approximately 1.2–1.5 pixels under substantially larger imposed displacements (up to ~140–200 pixels) and, uniquely, on speckle-free tissue surfaces. It is important to note that these supervised models are trained extensively on speckle-pattern datasets and within specific deformation regimes, which directly contributes to their high accuracy on such benchmarks. Our framework, by comparison, operates in a fully zero-shot manner without any retraining on bladder-specific images. Furthermore, the supervised DIC networks have not been evaluated under complex deformation modes such as active tissue contraction. Thus, while the absolute RMSE values from supervised DIC networks are slightly lower on standard speckled datasets, the present approach demonstrates greater flexibility for real biological applications where artificial speckling is undesirable or infeasible.

By design, our CoTracker3-based framework does not rely solely on pixel intensity correlation. Instead, it leverages convolutional feature extraction together with spatial and temporal attention mechanisms, enabling robust tracking of tissue features even through substantial deformation, folding, and buckling. The effectiveness of CoTracker3 is largely attributable to its sophisticated transformer architecture. Unlike conventional methods, transformers excel at capturing long-range dependencies and contextual relationships within sequential data. Specifically, the CoTracker3 model integrates two distinct attention mechanisms: spatial and temporal attention. Spatial attention employs the proxy tokens technique, which significantly reduces computational costs while maintaining the capability to track an extensive number of grid points (up to approximately 50,000). Temporal attention, meanwhile, enhances the model's robustness by effectively correlating points between successive frames, thus reliably maintaining tracking stability even amid dynamic and complex deformation sequences inherent in active bladder contractions.

Our custom isotonic biaxial contraction apparatus enabled mechanically and physiologically relevant conditions during testing. The bladder tissue deformed under biaxial loading conditions, approximating the biomechanical environment during active physiological contractions. Experiments with this system revealed distinct anisotropic strain distributions in the bladder, with statistically higher contraction in the longitudinal direction compared to the circumferential direction. Such anisotropy may have evolved in the ellipsoidal anatomical geometry of the rat bladder to optimize mechanical efficiency during urine expulsion. Although no previous study has directly examined the local strain field during bladder contraction, our findings are consistent with prior work reporting greater maximal longitudinal active stress generation in porcine bladder strips compared to the circumferential direction, attributed to greater longitudinal smooth muscle content [24]. Similarly, studies in rat bladder have demonstrated functional heterogeneity between longitudinal and transverse detrusor contractility, with longitudinal strips showing higher sensitivity to $K^+$ channel blockade, further supporting direction-dependent differences in bladder biomechanics [62].

Additionally, multiphoton microscopy (MPM) imaging complemented our quantitative findings by providing qualitative confirmation of the pronounced folding and buckling during contraction while also revealing structural features within the bladder. Our previous studies revealed that smooth muscle bundles have heterogenous orientations in the bladder [16]. Localized regions of elevated strain are observed in the present study, likely due to specific microstructural configurations of smooth muscle bundles and extracellular matrix distribution. In the future, these structural findings can be co-mapped to the quantitative strain fields to determine the structure function relationship between the organization of smooth muscles as the contractile elements and the deformation patterns during active contraction.

## 5- Limitations and Future Works

The method presented in this paper is based on single-camera video acquisition, and thus its application is currently limited to planar deformation analysis. While the multiphoton images revealed out-of-plane folding and buckling, the reported strains represent the in-plane projection of these deformations. In future studies, this framework can be extended to three-dimensional strain calculations using synchronized multi-camera systems, incorporating camera calibration and correspondence of identical grid points across image pairs. It is also worth noting that during bladder contraction, the sides of the folds may become trapped in valleys on the surface, so that surface features or speckles in these regions are not visible and cannot be tracked. Consequently, strain values in these regions must be interpolated between neighboring tracked grid points.

As with other DIC-based approaches, the density and quality of surface features influence the spatial resolution and accuracy of strain predictions. Although the chosen grid density provided stable estimates, systematic parametric studies of grid spacing and CoTracker3 hyperparameters could further optimize the trade-off between spatial resolution, noise sensitivity, and computational cost. Additional robustness tests under varying signal-to-noise ratios would help quantify sensitivity to imaging conditions.

While the consistency across the four biological replicates supports the finding of anisotropic contraction, future studies will expand the sample size to strengthen the statistical power for these scientific conclusions and explore the role of factors such as age and sex. In future studies, this framework will enable a mechanistic understanding of how voiding function is compromised for diseases such as Bladder Outlet Obstruction (BOO) by enabling a rigorous assessment of alterations to local strain fields during contraction.

## 6- Conclusion

We successfully developed and validated a novel speckle-free approach for experimental strain calculation in bladder tissue, utilizing the transformer-based CoTracker3 model. This framework enabled accurate and robust quantification and visualization of local strain fields during active bladder contraction without the need for artificial speckling or task-specific network training. We also introduced a compact isotonic biaxial apparatus that is portable, compatible with multiphoton microscopy, and allows the tissue to actively contract under physiologically relevant loading, in contrast to traditional stretch-controlled devices.

Across analytical benchmarks, a physical uniaxial tension experiment, and synthetic deformation of real bladder images, the method yielded low tracking (max error: 1.48 pixels) and strain errors, supporting its accuracy and robustness under large and complex deformations. Application to rat bladder contractions revealed heterogeneous and anisotropic strain distributions, with statistically greater longitudinal than circumferential shortening ($p<0.01$). These findings align with prior strip-level measurements of anisotropic detrusor contractility.

The non-invasive nature, broad applicability, and enhanced physiological relevance of this framework make it a valuable tool for advancing our understanding of soft tissue biomechanics in both healthy and pathological states.

**Funding:** The authors acknowledge the financial support provided by NIH grants: R01 AG056944 and R01 DK133434.

**Figures:**

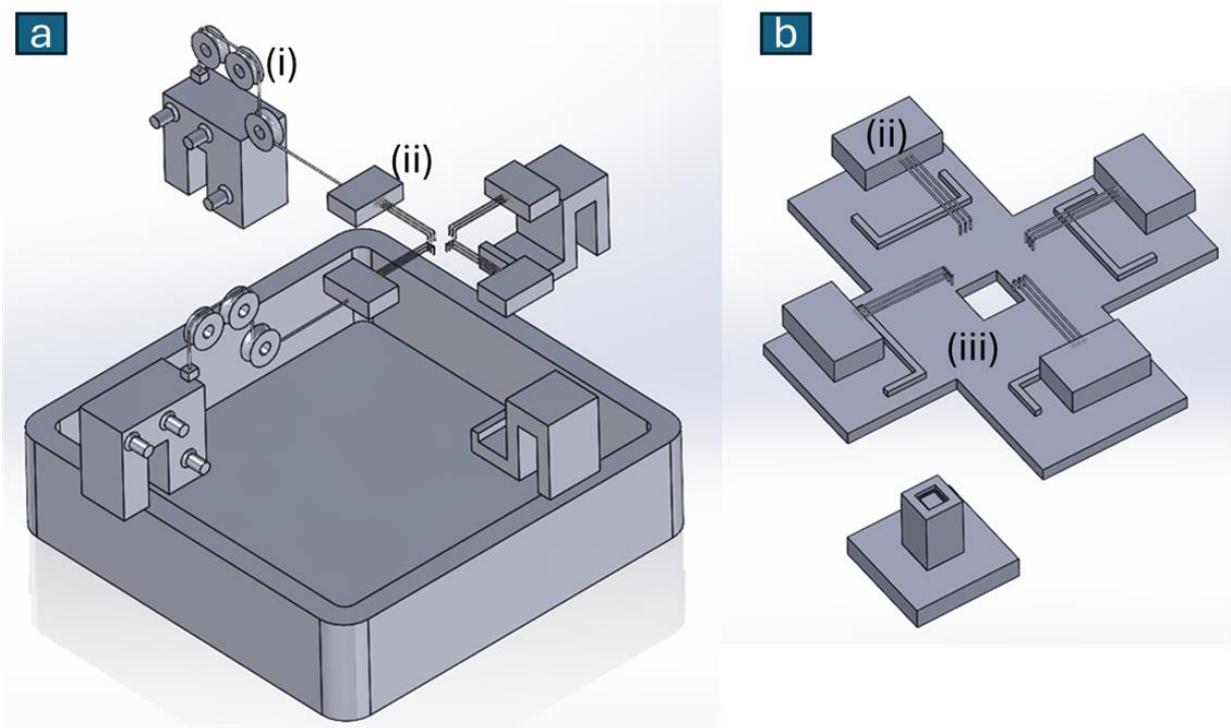

**Figure 1.** 3D models of the custom isotonic biaxial apparatus. (a) Exploded view of the apparatus, showing the solution tank and the triple-pulley loading mechanism (i) connected to the floating rake assemblies (ii). (b) Custom designed rakes (ii) and the sample mounting platform(iii).

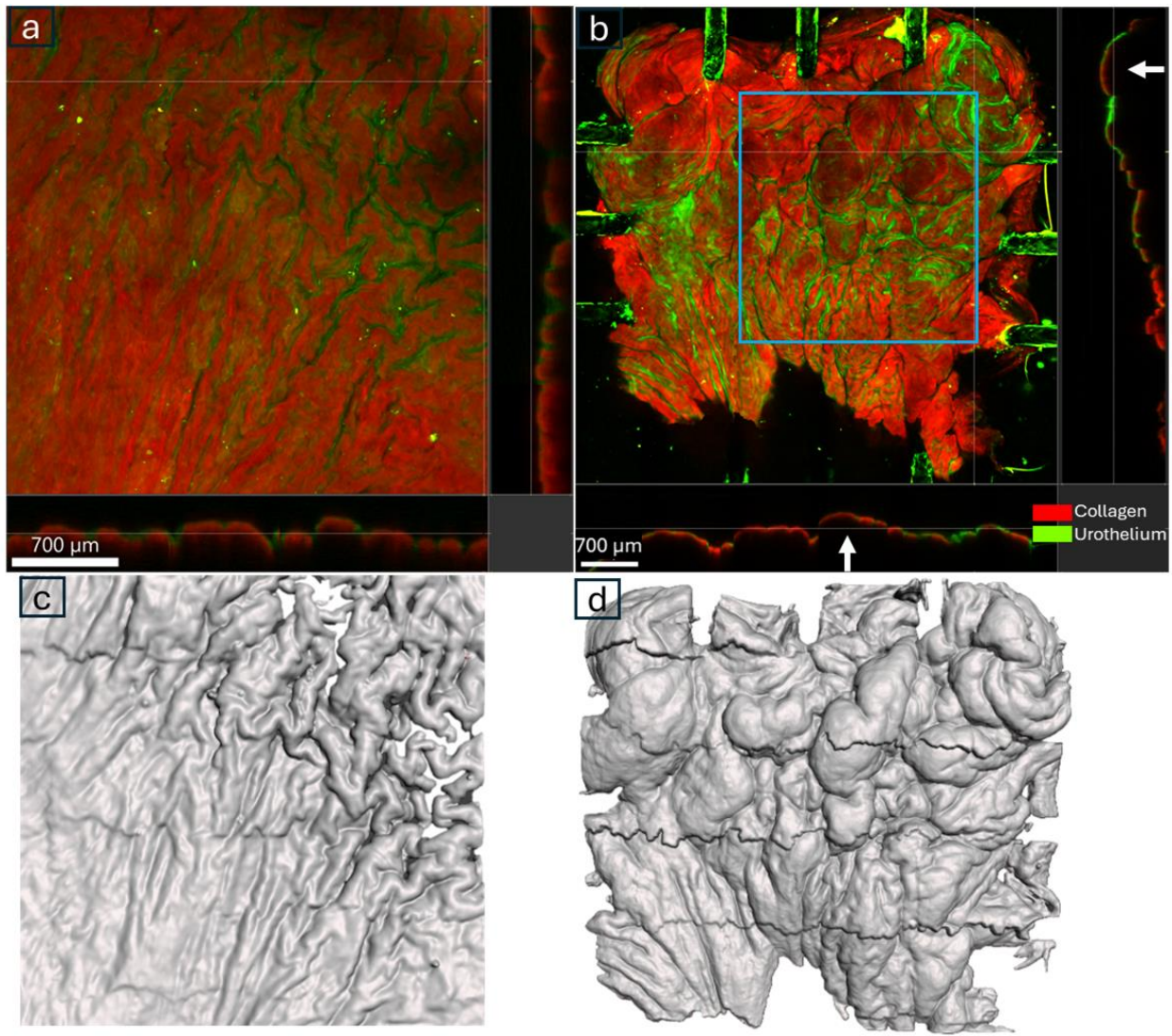

**Figure 2.** Multiphoton microscopy (MPM) images of bladder lumen before and after contraction. (a) MPM image of the bladder lumen before contraction. Small-amplitude undulations (rugae) of 50–150 µm are visible. (b) MPM image after contraction, showing large folds (maximum height ~ 900 µm) forming convex and concave surfaces. The blue square indicates the corresponding scanned area of figure a, in the contracted state, figure b. (c, d) 3D surface reconstructions corresponding to figures a and b, respectively.

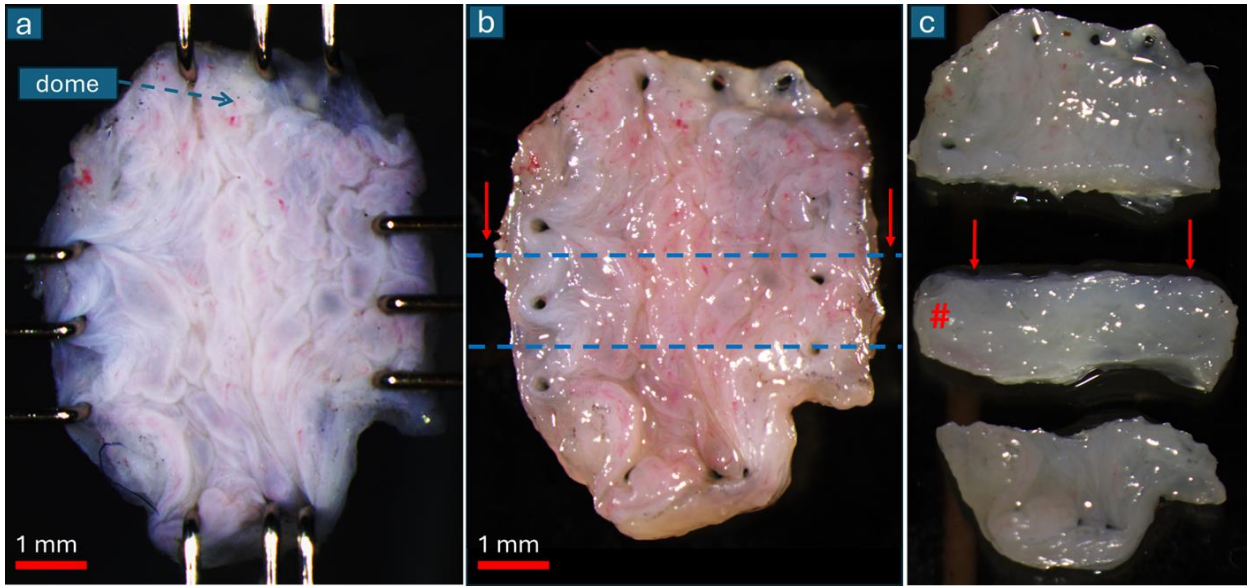

**Figure 3.** (a) Bladder in contracted state under the isotonic loading. (b) The sample was fixed while still being mounted on the device, by replacing the KREBS solution with PFA. The shape of the folds can be seen to be preserved. (c) a thick cross section of bladder (indicated by #) was cut from the central region (dashed line, Figure b), for MPM imaging of layer wise structure of the bladder wall. The scanned cross section is annotated by red arrows in Figures b and c.

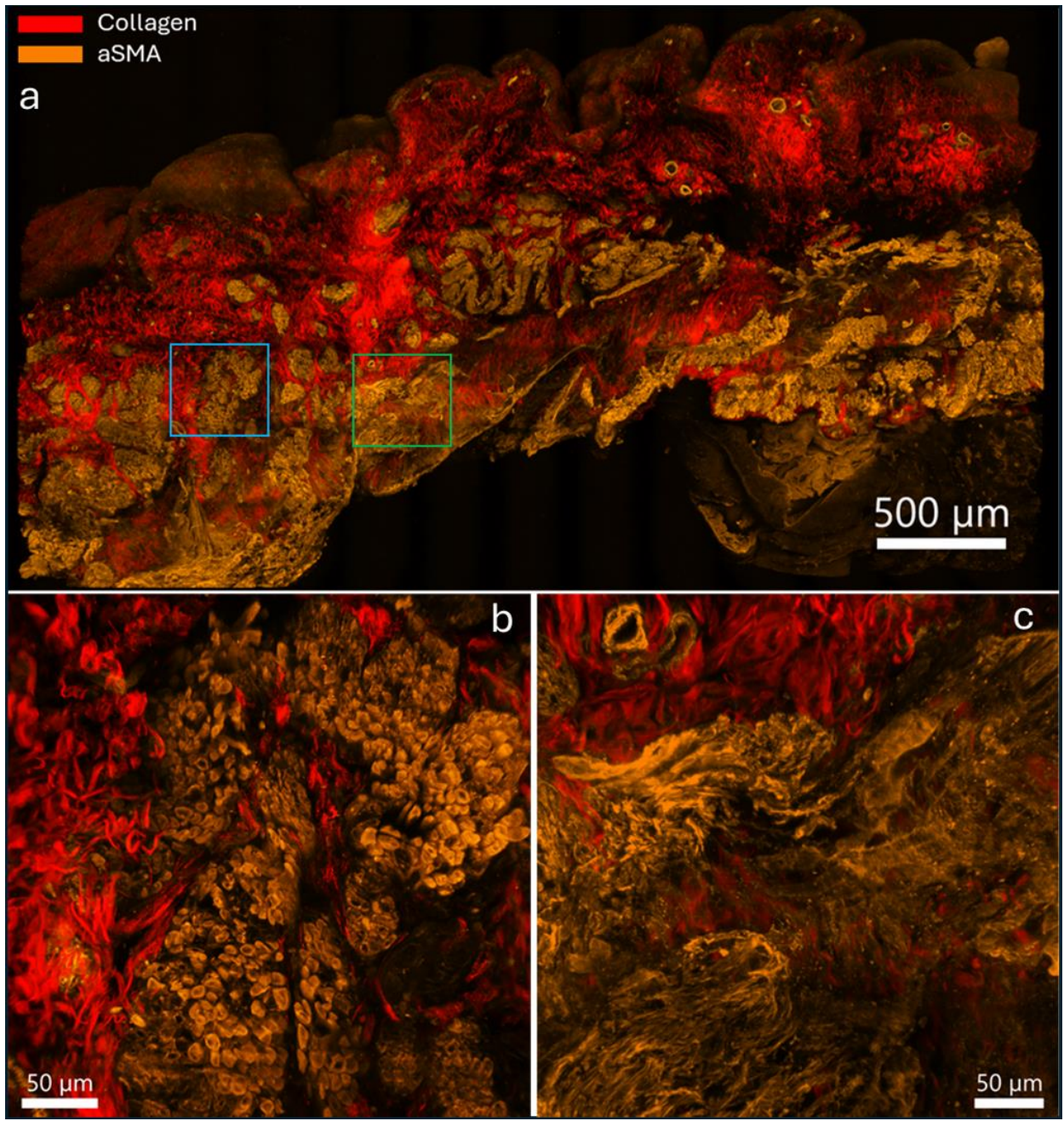

**Figure 4.** Layer-wise structure in contracted, fixed bladder. (a) MPM scan of a thick-cut section orthogonal to the lumen (lumen at top), showing autofluorescent collagen fibers (red) interwoven with alpha-smooth muscle actin-stained bundles (orange); large luminal folds are clearly observed. (b, c) Zoomed views of the regions indicated by blue and green squares in (a), respectively, highlighting out-of-plane and in-plane smooth muscle bundles, respectively. Transmural collagen fibers are observed, interwoven with the SMC layers.

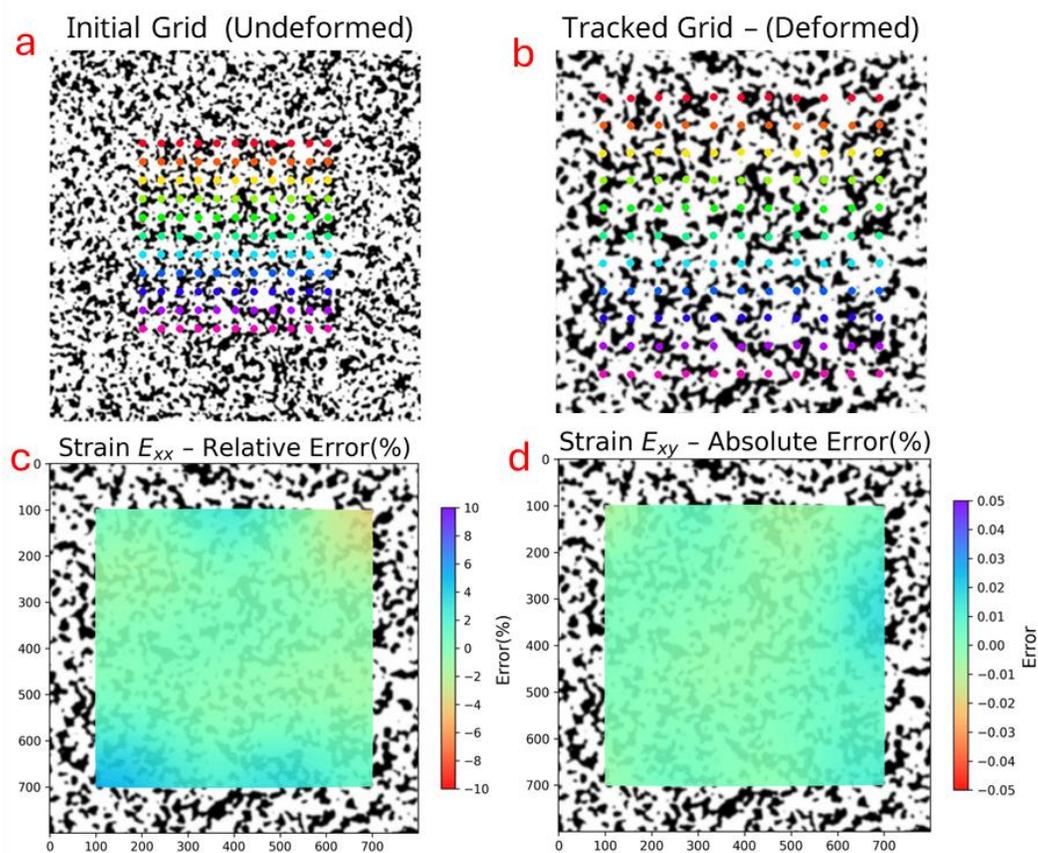

**Figure 5.** (a and b) Tracked grid points in a synthetic video of equibiaxial stretch. Synthetic speckles are shown in black, and colored dots indicate tracked points from the CoTracker3 model. The model achieved a tracking RMSE of 1.31 pixels (spatial SD: 0.62 pixels) on $800 \times 800$ resolution images. (c and d) Relative error (%) of strain components in the deformed state. The results show a strain RMSE of 0.0073 (SD: 0.0053) for $E_{xx}$ and 0.0047 (SD: 0.004) for $E_{xy}$, confirming that predicted strains closely match the analytical solution ($E_{xx} = E_{yy} = 0.625, E_{xy} = 0$).

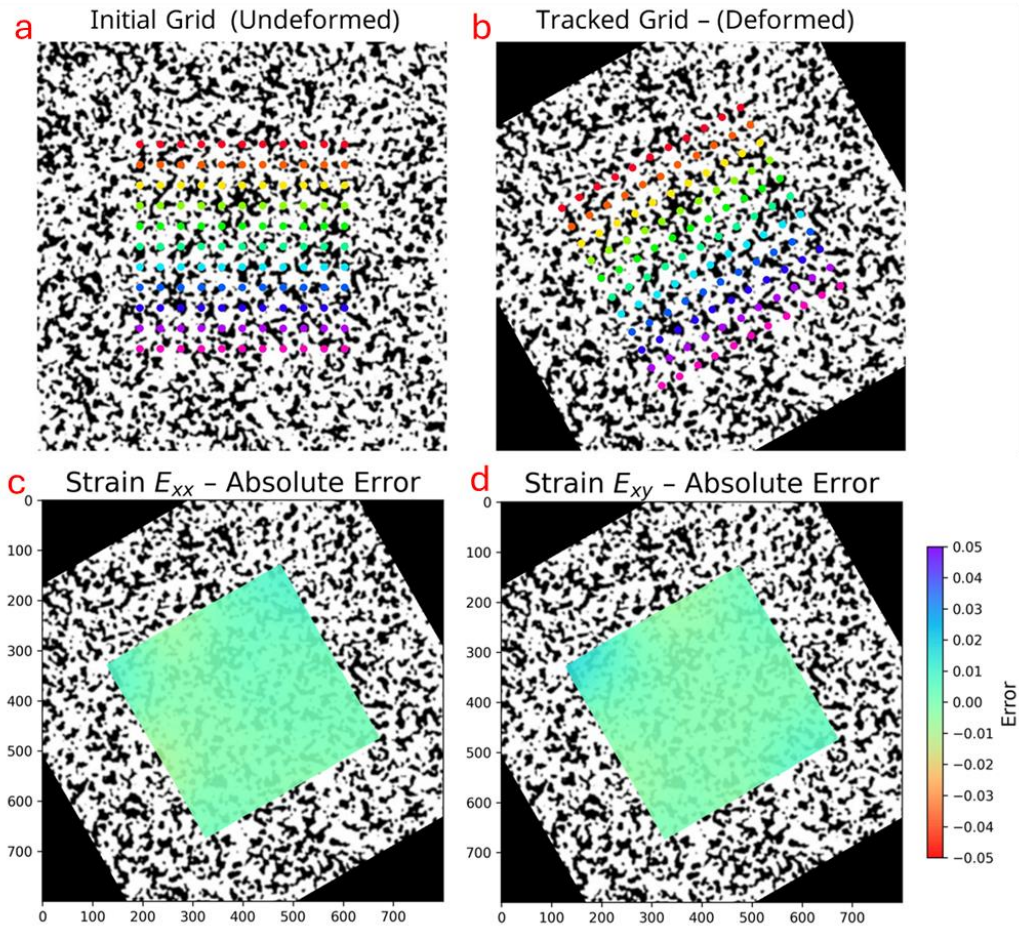

**Figure 6.** (a and b) Tracked grid points in a synthetic video of rigid body rotation (30° CCW). The model accurately tracked points with an RMSE of 1.29 pixels (SD: 0.64 pixels). (c and d) Contours of absolute strain error in the deformed state. The framework successfully preserved the zero-strain condition, yielding a strain RMSE of 0.0033 (SD: 0.0031) for $E_{xx}$ and 0.0037 (SD: 0.0035) for $E_{xy}$, consistent with the analytical solution ($E_{i,j} = 0$).

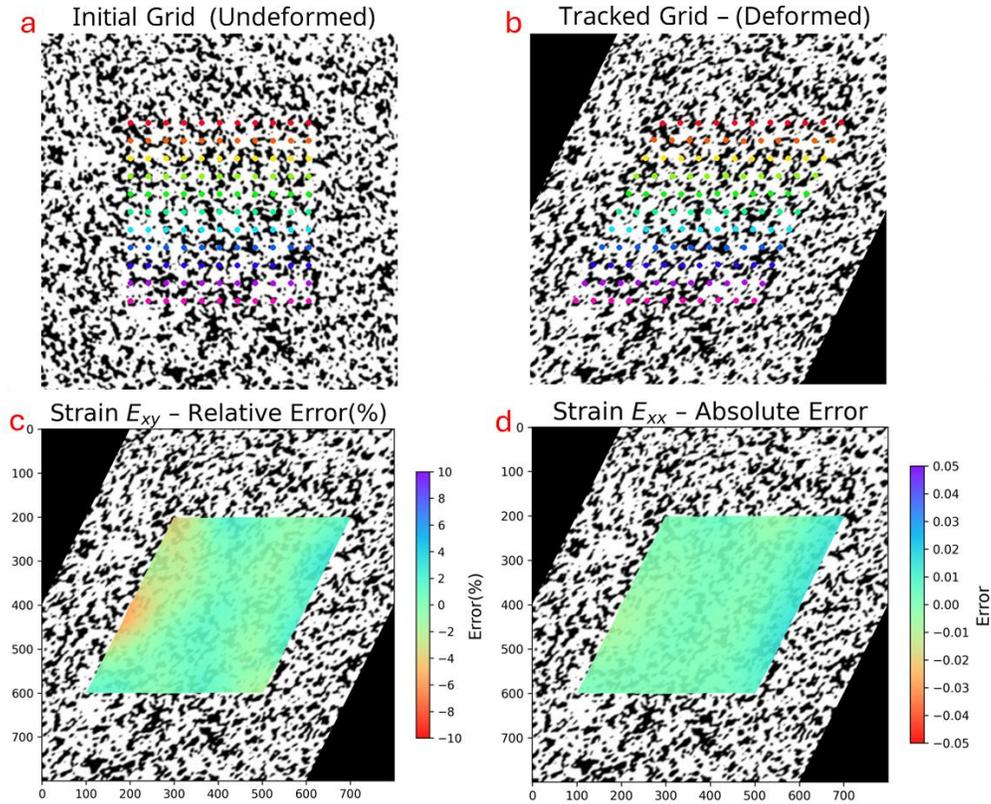

**Figure 7.** (a and b) Tracked grid points in a synthetic video of 2D shear deformation. The model achieved a tracking RMSE of 1.21 pixels (SD: 0.59 pixels) on $800 \times 800$ resolution images. (c and d) Contours of relative error for $E_{xy}$ and absolute error for $E_{xx}$ and in the deformed state. The results show a strain RMSE of 0.0045 (SD: 0.0041) for $E_{xx}$ and 0.0058 (SD: 0.0051) for $E_{xx}$ (absolute), closely matching the analytical solution ($E_{xx} = 0$, $E_{xy} = 0.25$).

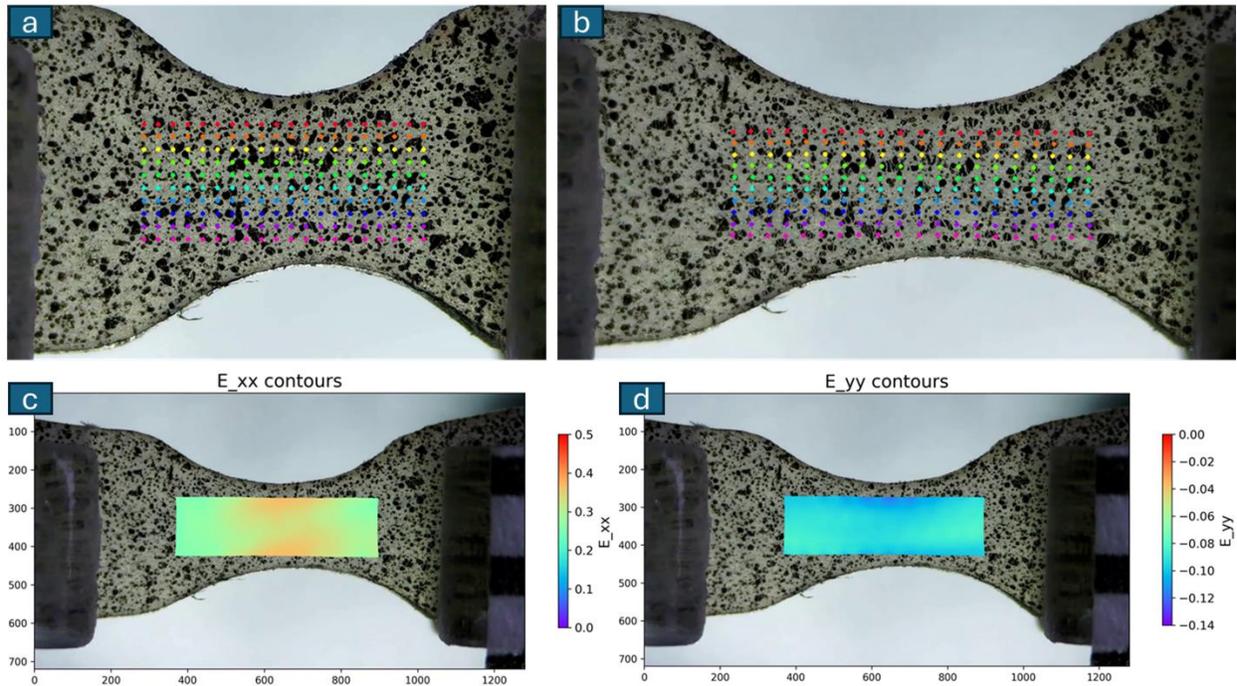

**Figure 8.** Tracked grid points in uniaxial stretch of the dog-bone shape latex sample before (a) and after deformation (b). The speckle pattern was applied using an airbrush to provide some texture on the inherently plain latex sheets. (c) and (d) show the strain components contours in the deformed state. Localized regions of elevated strain ($E_{xx}$) are observed in the narrower central region, which experiences higher stress due to the smaller cross-sectional area. Comparison of nodal predicted values with manual strain measurements (n = 5) indicated a strain RMSE of 0.0079 (SD: 0.0050) for $E_{xx}$ and 0.0130 (SD: 0.0100) for $E_{xx}$, validating the framework's performance in physical experimental conditions.

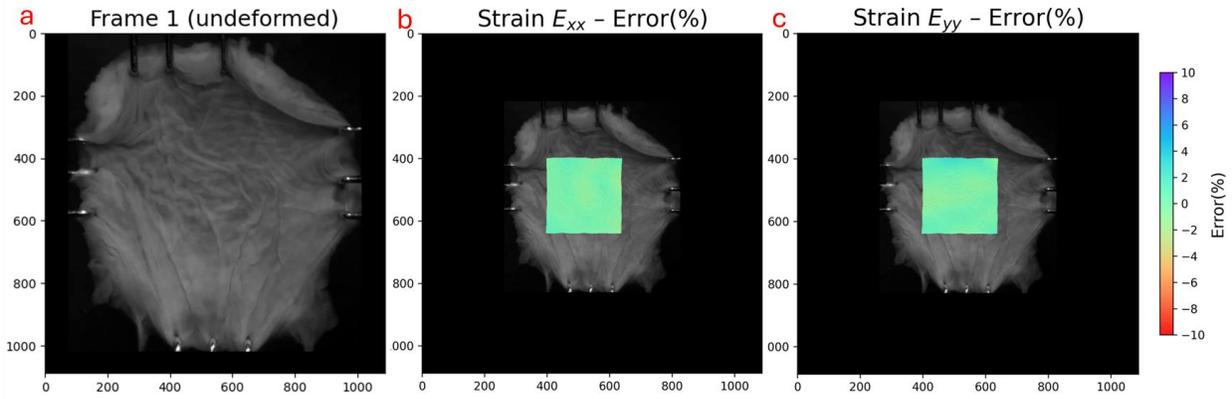

**Figure 9.** Synthetic deformation on natural bladder tissue texture. (a) Reference image of the bladder surface (undeformed frame). (b and c) Contours of relative error (%) for $E_{xx}$ and $E_{yy}$, respectively. The framework successfully recovered the imposed deformation on the natural, sparse texture of the bladder lumen, achieving a strain RMSE of 0.0083 (SD: 0.0076) for $E_{yy}$ and 0.0042 (SD: 0.0039) for $E_{xx}$. These predicted values closely match the ground-truth strains ($E_{yy} = E_{xx} = -0.32$).

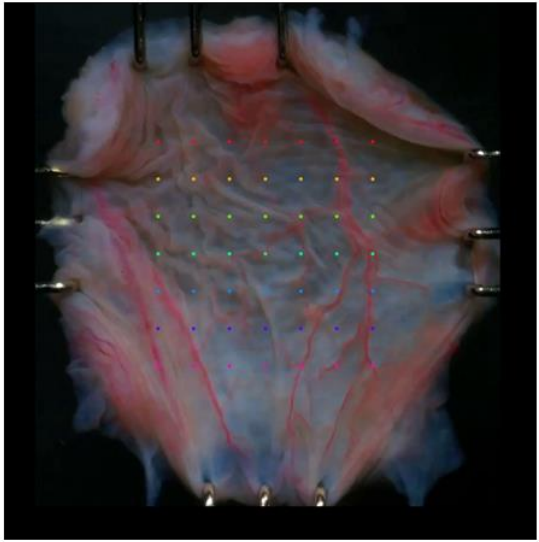
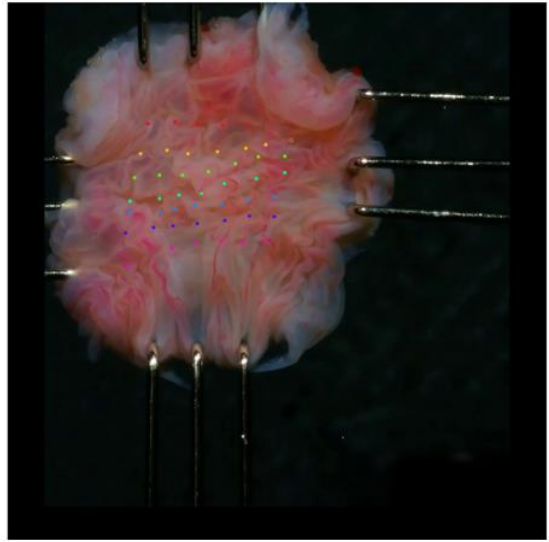
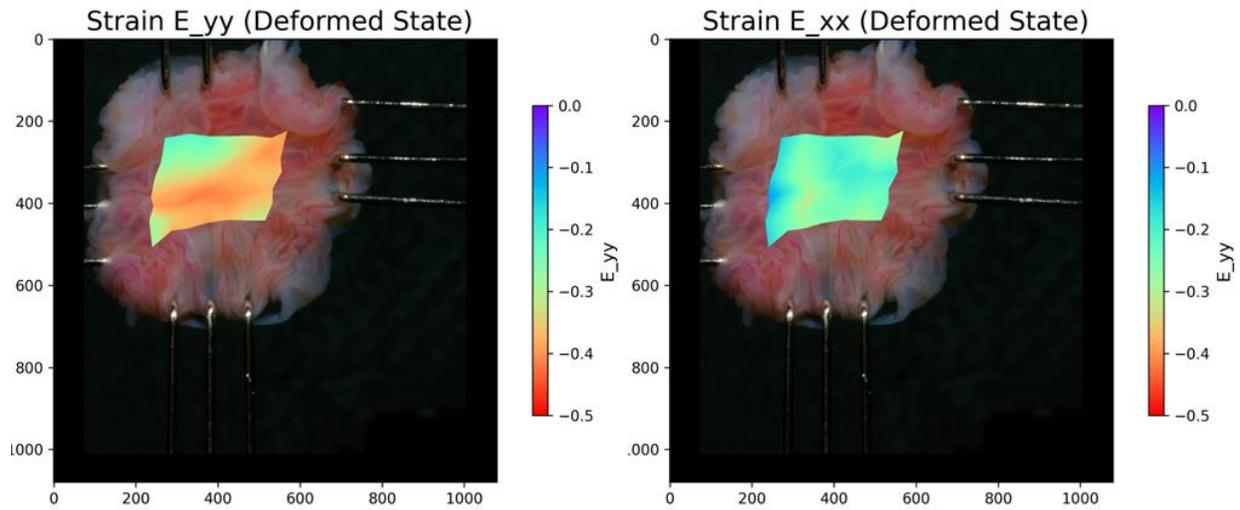

**Figure 10.** Sample 1 active contraction. (Top) Images of the bladder lumen before and after KCl-induced contraction with the virtual tracking grid overlaid. A 0.5 g weight was applied to each axis as the isotonic load. (Bottom) Calculated strain contours illustrating a heterogeneous and anisotropic local strain field.

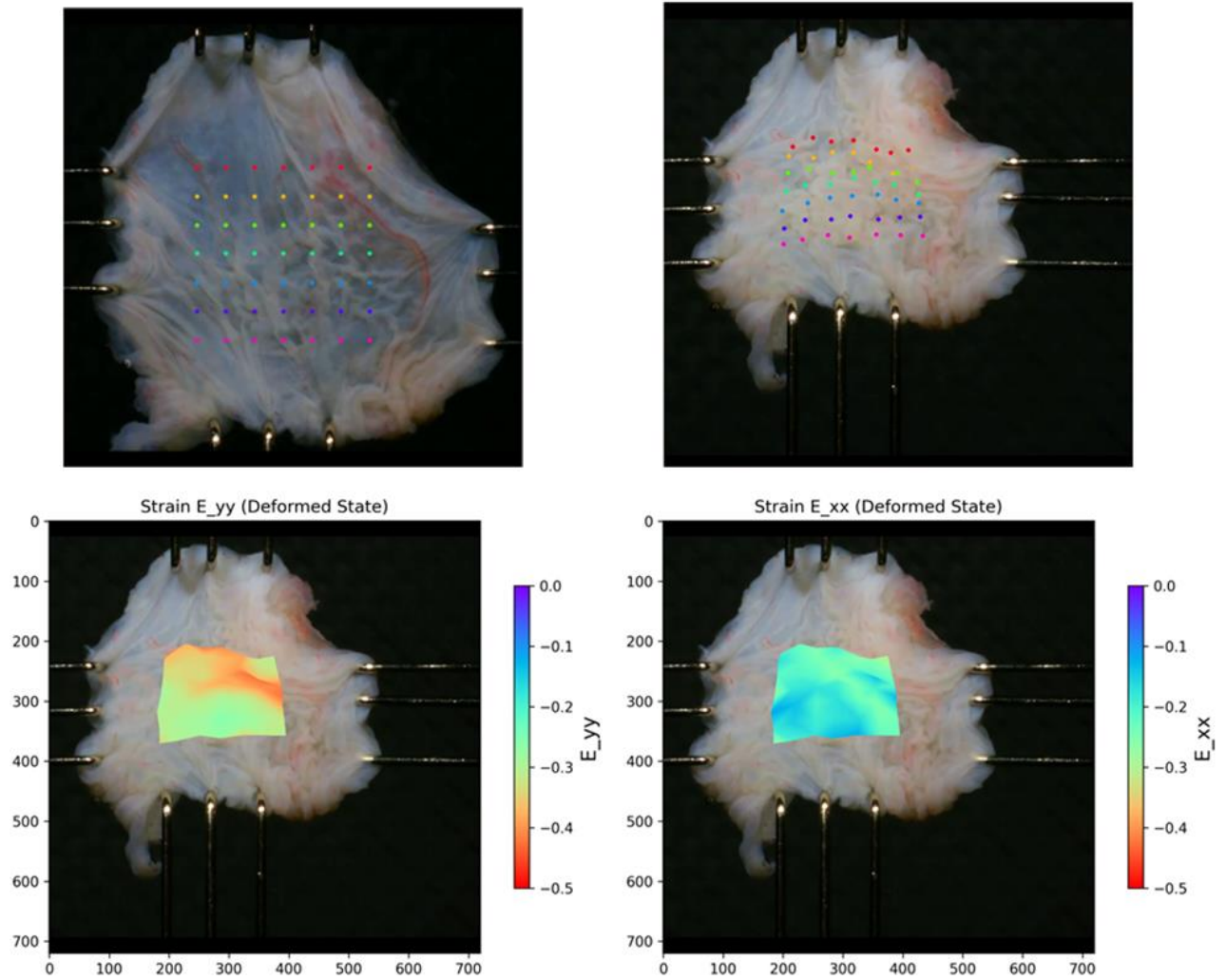

**Figure 11.** Sample 2 active contraction. (Top) Images of the bladder lumen before and after KCl-induced contraction with the virtual tracking grid overlaid. A 0.5 g weight was applied to each axis as the isotonic load. (Bottom) Calculated strain contours illustrating a heterogeneous and anisotropic local strain field.

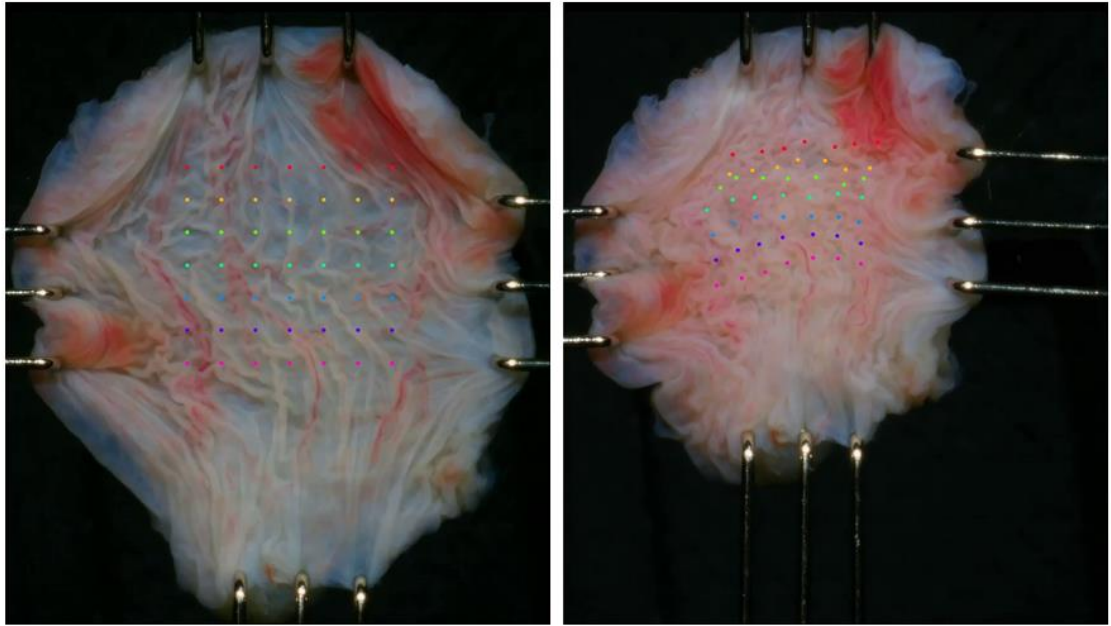
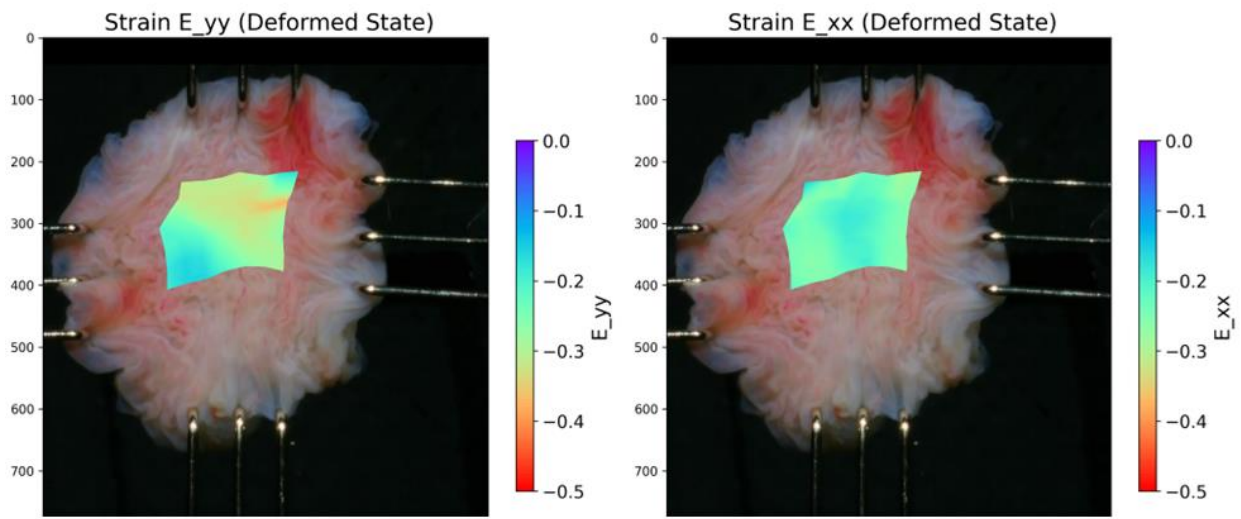

**Figure 12.** Sample 3 active contraction. (Top) Images of the bladder lumen before and after KCl-induced contraction with the virtual tracking grid overlaid. A 0.5 g weight was applied to each axis as the isotonic load. (Bottom) Calculated strain contours illustrating a heterogeneous and anisotropic local strain field.

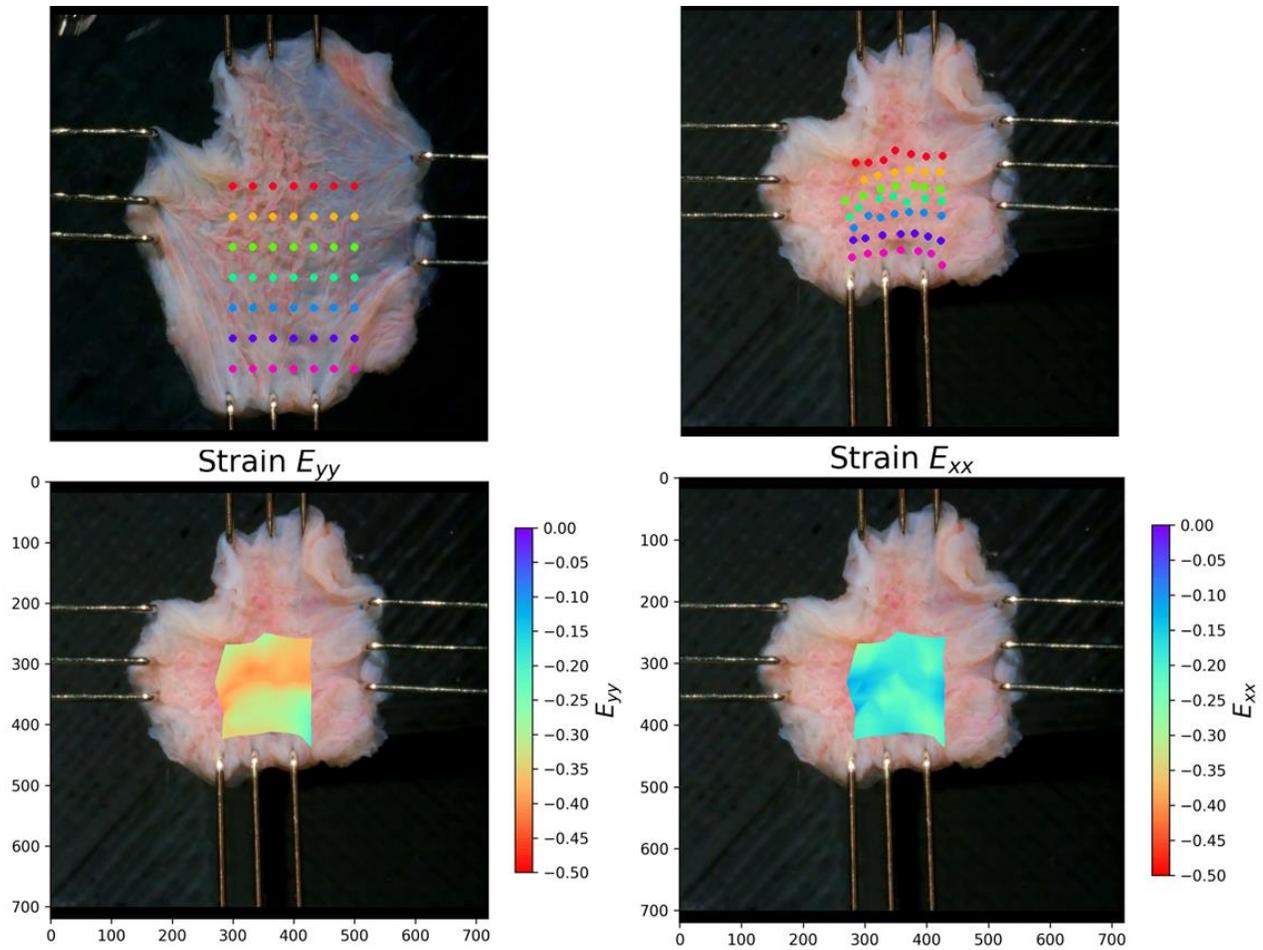

**Figure 13.** Sample 4 active contraction. (Top) Images of the bladder lumen before and after KCl-induced contraction with the virtual tracking grid overlaid. A 0.5 g weight was applied to each axis as the isotonic load. (Bottom) Calculated strain contours illustrating a heterogeneous and anisotropic local strain field

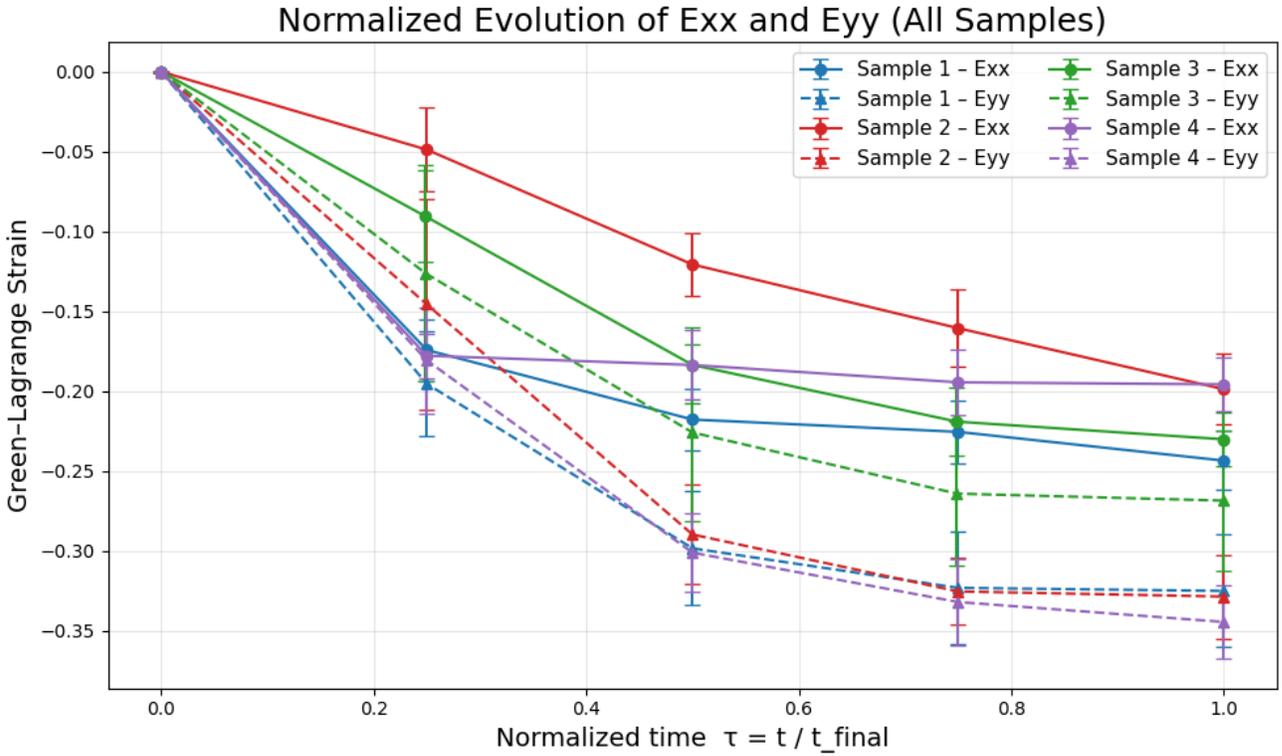

**Figure 14.** Temporal evolution of strain components. Evolution of mean Green-Lagrange strain components ($E_{xx}$ and $E_{yy}$) across all four specimens during active contraction. Both components increase monotonically, but the magnitude of $E_{yy}$ (longitudinal direction) consistently exceeds that of $E_{xx}$ (circumferential direction) across all samples, demonstrating significant contractile anisotropy. Error bars indicate the standard deviation, reflecting the heterogeneous nature of the deformation field.

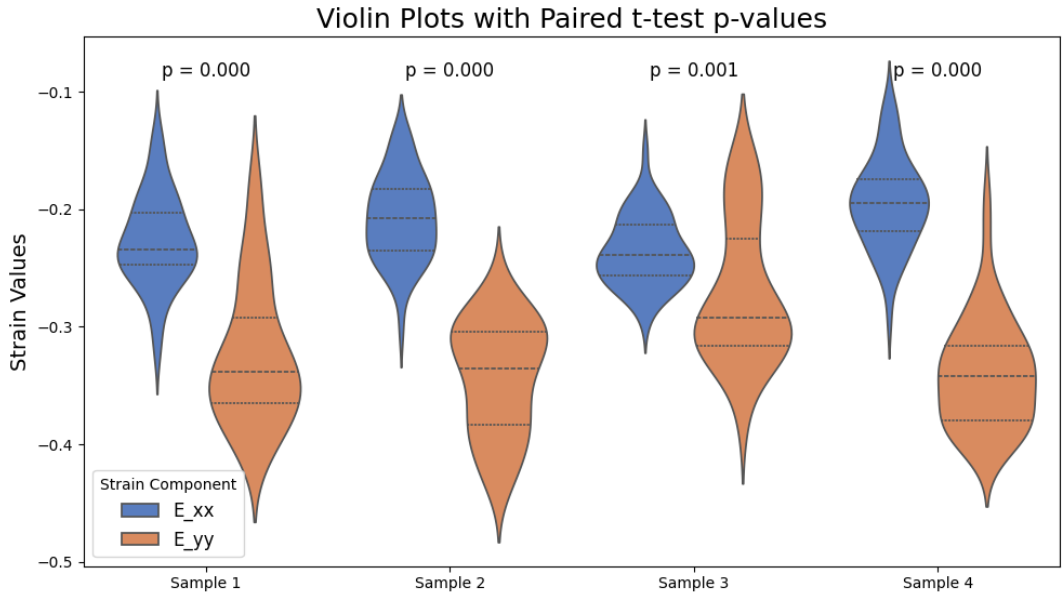

**Figure 15.** Statistical distribution of contractile strains. Violin plots showing the distribution of strain components for all specimens (n = 4). In all samples, the magnitude of longitudinal contraction ($E_{yy}$) was statistically significantly greater than the circumferential contraction ($E_{xx}$) ($p < 0.01$, paired t-test on n = 49 nodes per sample).